# The Impact of Planetary Rotation Rate on the Reflectance and Thermal Emission Spectrum of Terrestrial Exoplanets Around Sun-like Stars

Scott D. Guzewich[1,2] (301-286-1542), Jacob Lustig-Yaeger[3,4], Christopher Evan Davis[3,4], Ravi Kumar Kopparapu[1,2,4], Michael J. Way[5], Victoria S. Meadows[3,4]

[1]NASA Goddard Space Flight Center, 8800 Greenbelt Road, Greenbelt, MD 20771
[2]Sellers Exoplanet Environments Collaboration, NASA Goddard Space Flight Center, 8800 Greenbelt Road, Greenbelt, MD 20771
[3]University of Washington, Department of Astronomy, Box 951580, Seattle, WA 98195
[4]NASA NExSS Virtual Planetary Laboratory, Box 951580, Seattle, WA 98195
[5]NASA Goddard Institute for Space Studies, 2880 Broadway, New York, NY 10025

# ABSTRACT

Robust atmospheric and radiative transfer modeling will be required to properly interpret reflected light and thermal emission spectra of terrestrial exoplanets. This will help break observational degeneracies between the numerous atmospheric, planetary, and stellar factors that drive planetary climate. Here we simulate the climates of Earth-like worlds around the Sun with increasingly slow rotation periods, from Earth-like to fully Sun-synchronous, using the ROCKE-3D general circulation model. We then provide these results as input to the Spectral Planet Model (SPM), which employs the SMART radiative transfer model to simulate the spectra of a planet as it would be observed from a future space-based telescope. We find that the primary observable effects of slowing planetary rotation rate are the altered cloud distributions, altitudes, and opacities which subsequently drive many changes to the spectra by altering the absorption band depths of biologically-relevant gas species (e.g., $H_2O$, $O_2$, and $O_3$). We also identify a potentially diagnostic feature of synchronously rotating worlds in mid-infrared $H_2O$ absorption/emission lines.

# 1. INTRODUCTION

Recently, terrestrial-size or mass planets have been detected in the habitable zones of a variety of host stars, some of which include Proxima Centauri b [Anglada-Escudé et al., 2016], Kepler 62f [Borucki et al., 2013], Kepler 442b [Torres et al., 2015], Ross 128b [Bonfils et al., 2018], and the TRAPPIST-1 system [Gillon et al., 2017]. Detailed transmission spectroscopy across the near-infrared is within reach of the James Webb Space Telescope for numerous known terrestrial-sized exoplanets [Barstow and Irwin, 2016; Morley et al., 2017; Batalha et al., 2018; Lustig-Yaeger et al., 2019]. Furthermore, the Transiting Exoplanet Survey Satellite is expected to identify many new terrestrial planets in the habitable zones of nearby stars which will be excellent candidates for follow up observations [Ricker et al., 2015; Sullivan et al., 2015; Barclay et al., 2018]. Upcoming ground-based telescopes such as the European Extremely Large Telescope and the Thirty Meter Telescope should be able to directly image Proxima b [Turbet et al., 2016; Lovis et al., 2017]. Future space telescopes that are currently being studied, such as the Large Ultraviolet Optical and Infrared telescope (LUVOIR) [Fischer and Peterson et al., 2019], the Habitable Exoplanet Observer (HabEx) [Gaudi and Seager et al., 2019], and the Origins Space Telescope (OST) [Cooray et al., 2019], are being specifically designed for characterizing the atmospheres of terrestrial extrasolar planets in the habitable zone. In the next decade, retrieving spectra from terrestrial planets in the habitable zones of low-mass stars may be within reach. Within the next two decades, we may be able to retrieve spectra from terrestrial planets in the habitable zones of Sun-like stars [Fischer and Peterson et al., 2019; Gaudi and Seager et al., 2019]. The day is close at hand where we will retrieve high quality observations of habitable extrasolar worlds around G, K, and M stellar spectral types. However, our observations will only be as good as the models used to interpret them.

Terrestrial planets are small and significantly more challenging to observe compared to their larger gaseous brethren. While there is a possibility that observatories coming online in the coming decade will begin to allow their characterization, it is not enough to invest only in observational endeavors. Comprehensive modeling of planetary atmospheres is required in order to fully understand the effect on observables. 3D general circulation models (GCMs) allow for coupled, self-consistent, multidimensional simulations, which can realistically simulate the plausible climates of terrestrial extrasolar planets. A complete theoretical understanding of terrestrial exoplanetary atmospheres, gained through comprehensive 3D modeling, is critical for interpreting spectra of exoplanets taken from current and planned instruments, and is critical for designing future missions that aim to measure spectra of potentially habitable exoplanets as one of their key science goals.

In the past decade, GCMs have become commonly used to study terrestrial planets in or near the habitable zone (a non-exhaustive list includes: Wordsworth et al., 2011; Abe et al., 2011; Pierrehumbert, 2011; Selsis et al., 2011; Leconte et al., 2013; Wang et al.,. 2014; Yang et al., 2014a, 2014b; Shields et al., 2013, 2016; Koll & Abbot, 2015; Wordsworth et al., 2015; Pierrehumbert & Ding, 2016; Bolmont et al., 2016; Way et al., 2016, 2017, 2018; Popp et al., 2016; Wolf et al., 2017; Kopparapu et al., 2017; Turbet et al., 2016; 2017; Boutle et al., 2017; Fujii et al., 2017; HaqqMisra et al., 2018; Kodama et al., 2018; Del Genio et al., 2019; Jansen et al., 2019; Yang et al., 2019a). However, the climate of planets depend on many parameters, including the total stellar irradiation, the atmospheric composition, the volatile inventory, the planetary rotation rate, the surface properties, and the presence of clouds or hazes [Forget & Leconte, 2014]. While exploring all these parameters is beyond the scope of this study, we focus on varying the rotation rate of an Earth-like planet around a Sun-like star to study its effect on the reflectance and thermal spectra, which is relevant for direct imaging observations. The motivation arises from considering the diversity of current exoplanet discoveries with varied characteristics. While mission studies like LUVOIR and HabEx rightly focus on Earth-analogs, with Earth-like characteristics (ex: Earth's mass/size, atmosphere, rotation) to constrain their mission capability and instrument design performance, we should also expect that it may be unlikely that we may find a true Earth-analog. What features in the observed spectrum are affected when one deviates from Earth-like assumptions? What can we infer about the climate of the planet if we have time series, phase-dependent observations of an "Earth" cousin with different physical and atmospheric characteristics? Are habitable conditions possible on such a planet? Our goal in this paper is to explore a small part of this question, by varying just the rotation period of the Earth, while keeping other factors similar to Earth.

In fact, such a study has been considered by Jansen et al. [2019] using the same GCM (Resolving Orbital and Climate Keys of Earth and Extraterrestrial Environments with Dynamics, ROCKE-3D, Way et al. 2017) that we use in this study. They varied the rotation period of an Earth-like planet from 1x to 256x the sidereal day length of present Earth. Interestingly, they find that the fractional habitability, which they defined as the fractional surface area that maintains a mean temperature between 0°C and 100°C, reaching a maximum around

16x to 32x rotation period (a multiple of modern Earth's sidereal day length) depending upon the incident stellar flux on the planet. As discussed in the first paper of their series [Way et al. 2018], this is due to the transition between rotational regimes, where the slow rotation (1x to 16x) initially increases the fractional habitability by expansion of Hadley cells to higher latitudes, thereby transporting heat from warm, equator regions to cold, polar regions. However, beyond 16x to 32x rotation periods, the day length is longer than the radiative relaxation time of the atmosphere, and contrast in the day-to-night temperatures increases significantly. Furthermore, it is also possible that beyond 16x to 32x rotation periods, the efficacy of poleward heat transport through Hadley cells is maximized, and substellar cloud reflection dominates by increasing the albedo, cooling the planet and in turn reducing the fractional habitability. Jansen et al. [2019] also find that a similar peak is observed for carbonate-silicate weathering process, but around ~4x rotation period because of precipitation and temperature changes from rotational evolution. These effects illustrate the importance of varying rotation rate on the habitability of an Earth-like planet. In our study, we discuss the observational implications of the climate states of the planet, which was not discussed in Jansen et al. [2019]. Note that while the simulations used by Jansen et al. [2019] are very similar to those discussed below, they are not identical.

The structure of the paper is as follows: In Section 2 we describe the GCM and line-by-line radiative transfer models used in this work. In Section 3 we present our climate and disk-integrated planetary spectrum results. In Section 4 we discuss the implications and significance of our findings, and we conclude in Section 5.

## 2. MODEL DESCRIPTIONS

The modeling approach of our study is to simulate Earth with a variety of different slow rotation rates using a GCM, and then post-process the GCM climate results with a line-by-line radiative transfer model to study the observables that may be linked with slowly rotating Earth-like exoplanets. In the next subsections, we describe the ROCKE-3D GCM and the Spectral Mapping Atmospheric Radiative Transfer (SMART) model, which is the core of the SPM.

### 2.1. ROCKE-3D GCM

ROCKE-3D is the generalization and adaptation of the NASA Goddard Institute for Space Studies (GISS) ModelE2 GCM to planetary and exoplanetary worlds [Way et al., 2017]. ModelE2 is a terrestrial climate model currently being used for the Coupled Model Intercomparison Project 5 and 6 (CMIP5, CMIP6) [Schmidt et al., 2014]. Much of the ROCKE-3D model's code, including the dynamical core and many of the physical parameterizations, are identical between the parent terrestrial GCM and the modified planetary/exoplanetary version, termed "Planet 1.0", that we employed for these simulations. The model is freely available to the public and can be downloaded from the NASA GISS website (https://simplex.giss.nasa.gov/gcm/ROCKE-3D/). Amongst other ongoing work, ROCKE-3D has been used previously to simulate hypothetical ancient Venus configurations to study the inner edge of the habitable zone [Way et al., 2016], the habitability of Proxima Centauri b [Del Genio et al., 2018], and the moist greenhouse limit [Fujii et al., 2017].

While ROCKE-3D derives the bulk of its code heritage from the GISS ModelE2 GCM, some changes were required to make the code adaptable to, e.g., changing orbital parameters, different stellar types and insolation rates, varying atmospheric compositions and pressures, etc. The largest of these changes was through the incorporation of the Suite of Community Radiative Transfer codes (SOCRATES) radiation scheme [e.g., Amundsen et al., 2016]. SOCRATES is flexible to insolation and gaseous absorption/scattering variations from the terrestrial baseline to known planetary (e.g., Mars) and hypothetical exoplanet environments. SOCRATES employs a correlated-k method for gaseous absorption in the longwave and shortwave streams [Edwards 1996; Edwards and Slingo, 1996]. The number of bands chosen in each stream can be changed based on the required precision of the radiative transfer, stellar spectrum, and abundances of various radiatively-active gas species in the simulated atmosphere.

Our simulations are investigating the variability in the reflectance and thermal emission spectra due to rotation rate alone, so we limited the (potentially very large) parameter space that we explored to that factor in exclusion of others (e.g., such as surface land/water distributions, atmospheric composition, etc.). All simulations were run with 4° latitudinal and 5° longitudinal resolution, 40 atmospheric vertical layers spanning the surface (~1000 mb) to 0.1 mb, and 13 layers in a fully-coupled dynamical ocean with depths reaching ~5000m in places. To prepare each simulation, we altered the sidereal rotation period (i.e., sidereal day length) in the model's "rundeck". The "rundeck" is simply the input file that prepares a given simulation's parameters. Our rundecks are archived at Zenodo (https://doi.org/10.5281/zenodo.3727183). Table 1 specifically defines the sidereal rotation period of each simulation in seconds and the names of each simulation that we will use for the remainder of this paper. For comparison, Earth's sidereal rotation period is 86164.094 seconds and sidereal year is 31558150.23 seconds. Our focus in this work is on slower rotation periods and G-type stars as rotation rate (specifically tidal locking or resonant rotation) has been discussed thoroughly in the literature especially relating to M-type stars and their planets [e.g., Kopparapu et al., 2017; Wolf, 2017] and the inherent limitations of the GCM at faster rotation rates. While the ROCKE-3D GCM can simulate modestly faster rotation and maintain stability (e.g., a sidereal day length ½ that of Earth's), there is more rotation rate parameter space that can be simulated by the GCM at slower rotation rates. Additionally, using the examples provided by nature in our own Solar System, there is a bias toward slower rotation amongst the terrestrial planets with even best estimates of early Earth's day length only slightly shorter than the modern value (i.e., ~21-22 hours [Williams, 2004]). In addition, work by Barnes [2017] has shown that if Earth had an initial rotation period of ~3 days and never experienced a moon impactor it may have reached synchronous rotation after ~4.5 Gyr, hence spending much of its spin evolution in the slowly rotating regime (period > 16 sidereal days). Although atmospheric tidal forcing may continue to drive asynchronous slow rotation (e.g., Leconte et al., 2015).

Table 1. List of simulations.

| Name | Sidereal Rotation Period (seconds) | Obliquity |
|------|-----------------------------------|-----------|
| 1x   | 86164.094                         | 23.5°     |
| 4x   | 344656.38                         | 23.5°     |
| 8x   | 689312.75                         | 23.5°     |
| 16x  | 1378625.5                         | 23.5°     |
| 32x  | 2757251.0                         | 23.5°     |
| 64x  | 5514502.0                         | 23.5°     |
| 128x | 11029004.0 | 23.5° |
| 243x | 21038480.0 | 23.5° |
| 256x | 22058008.0 | 23.5° |
| 365x | 31558150.23 | 23.5° |
| 365x0° | 31558150.23 | 0° |

Other customizable factors in the simulation were left as Earth-like as possible. The background atmosphere was identical to Earth's with $CO_2$ present at 400 ppm, and $O_3$ and $CH_4$ present in their default terrestrial configurations that exhibit a prescribed slight seasonal variation in abundances with latitude and pressure. The input stellar spectrum was the modern solar spectrum and the remainder of Earth's orbital parameters were retained except for the "365x0°" simulation where obliquity was set to 0° to simulate a truly tidally-locked world. The topographic map was slightly altered from the terrestrial baseline by eliminating most islands (e.g., Japan, New Zealand, New Guinea, the Caribbean islands, etc.) and closing narrow sea passages (e.g., the Mediterranean, Red Sea, Persian Gulf, Hudson Bay). All vegetation was eliminated from the land and the surface soil was set to be a 50/50 mix of sand and clay with a set albedo of 0.3. Ocean bathymetry was simplified to a "bathtub" ideal with a set 1360 m ocean depth and limited ocean shelves. This prevents the ocean freezing to the bottom in shallow seas, which crashes the simulation. See Way et al. [2018] for more details on these modifications which are the same as those used herein. The model self-consistently models the water cycle to include dynamically variable lakes and snow and ice coverage. We excluded all aerosols (e.g., dust, black carbon, or volcanic aerosols) from the simulations.

Each simulation was run until it reached radiative equilibrium, defined as being less than +/- 0.2 W/m², and then simulated for an additional 100 years (Figure 1). For the 243x, 256x, and 365x simulations, the resonant, near-resonant, or synchronous rotation rates (respectively) of the planets introduce semi-regular multi-decadal variations in the global net radiative balance (Figure 1, bottom row). These runs were evaluated as being in radiative equilibrium over the long term by simulating for an additional length of time and having consistent and stable global mean surface temperatures. Except where specified, the data shown below is an average of the last 10 years of each simulation.

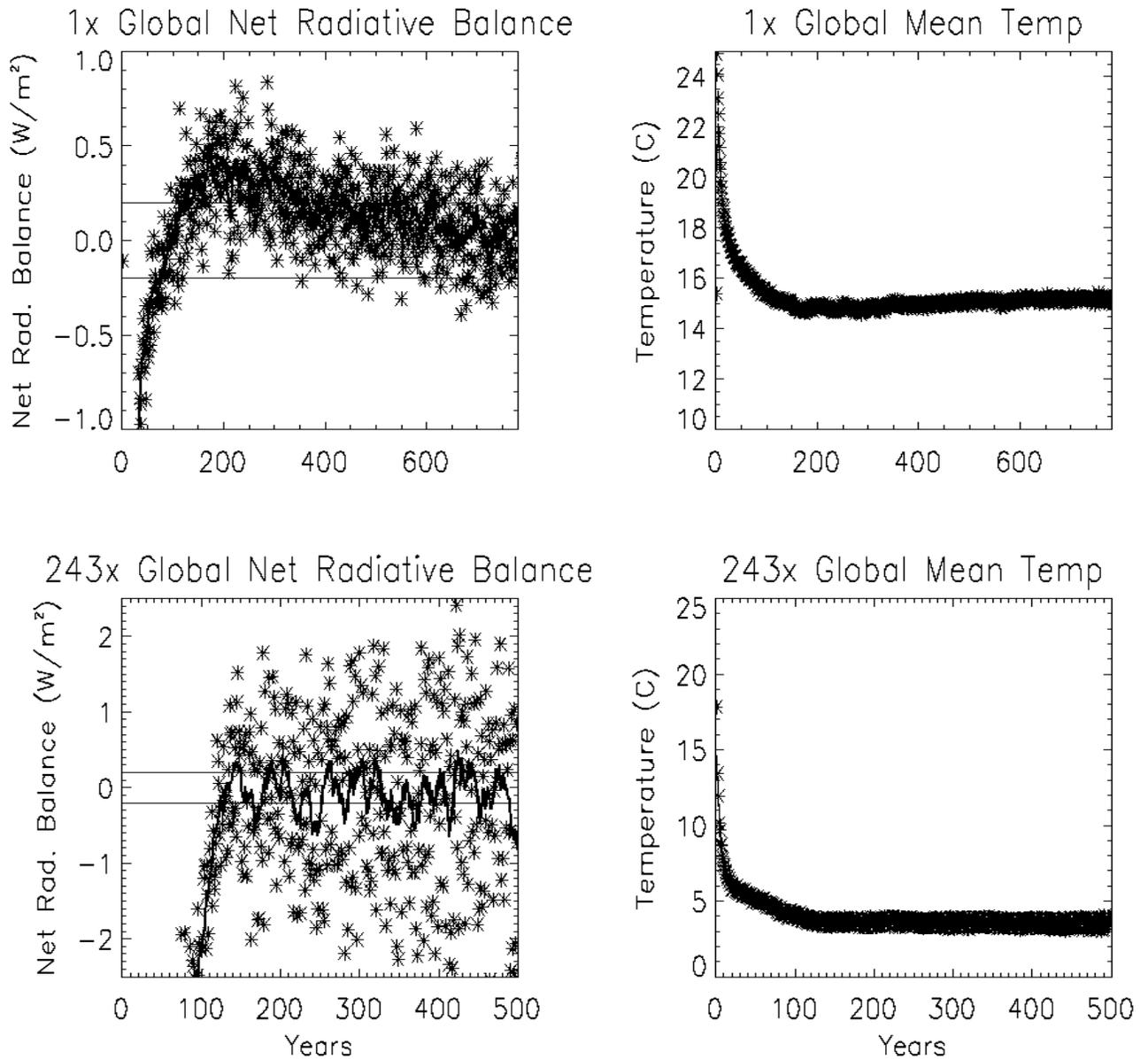

Figure 1. Global net radiative balance (W/m², left column) and global mean surface temperature (°C, right column) over simulated years for the 1x (top row) and 243x (bottom row) simulations. The asterisks are individual year values and the thick black line is the 10-year running mean. The thin black lines on the left column plots delineate the +/- 0.2 W/m² bounds.

2.2. SMART

To investigate the spectroscopic observables simulated in this work that may not be present in the lower resolution radiative transfer that is native to ROCKE-3D, we developed an interface between the ROCKE-3D GCM and the Spectral Mapping Atmospheric Radiative Transfer (SMART) model [Meadows and Crisp, 1996]. This model builds upon and generalizes the Virtual Planetary Laboratory (VPL) 3D Spectral Earth Model [Robinson et al., 2010, 2011, 2014] to allow for the spectroscopic visualization and disk-integration of planets described by GCM output states. We refer to this new code as the VPL Spectral Planet Model (SPM), which has three coupled modules that are used in series to (1) interface the essential ROCKE-3D output variables with SMART input variables; (2) simulate the atmospheric radiative transfer with SMART for each resolved planet pixel; and (3) integrate the visible pixels to create disk-integrated spectra. We describe these three modules below.

2.2.1. ROCKE-3D to SMART Translation

We interface ROCKE-3D with SMART by translating various parameters from the GCM's output files into files readable by SMART. These parameters include surface temperature, atmospheric composition as a function of pressure, atmospheric temperature structure, liquid water and ice cloud optical depths as a function of pressure, cloud covering fractions, and ground albedos that are independent of wavelength (i.e. "grey"). These quantities are binned into Hierarchical Equal Area isoLatitude Pixelization (HEALpix) [Gorski et al., 2005] pixels, weighting the contribution by the GCM's latitude-longitude pixel area, which varies with latitude. We employ the HEALPix scheme, which covers the planet globe with equal area pixels, to avoid the geometrical instabilities, such as diminishing pixel size and increasingly triangular pixel shape, that arise near the poles of a latitude-longitude grid and to simplify our spectral disk-integration calculations. Our nominal simulations are run with NSIDE, a parameter that defines the resolution of a HEALPix grid, equal to 2 for a total of 48 pixels across the planet globe. An example of a HEALPix grid with this resolution is shown in Figure 2. The final binned quantities are written to ASCII files that are used as inputs to SMART. Note that the module to interface ROCKE-3D with SMART can be swapped out at a later time with interfaces for other GCMs, to expand the functionality of this model over time.

2.2.2. Running SMART and Database Creation

After translation, we run atmospheric radiative transfer with SMART for each HEALpix planet pixel. SMART solves the equation of radiative transfer in plane-parallel geometry using line-by-line, multi-stream, multi-scattering, solar and thermal calculations for vertically-resolved atmospheric structures. SMART was developed for Solar System terrestrial planet atmospheres and has been rigorously validated on Earth [Robinson et al., 2011] and Venus [Meadows and Crisp, 1996; Arney et al., 2014], as well as applied to study a diversity of exoplanet environments, including the spectra of M-dwarf habitable planets [Segura et al., 2003, 2005; Meadows et al., 2018; Lincowski et al., 2018], habitable haze-enshrouded exoplanets [Arney et al., 2016; 2018] and $H_2$-dominated mini-Neptunes [Charnay et al., 2015]. Water cloud wavelength-dependent and scattering optical properties used by SMART are specified by phase as either cirrus clouds (liquid) [Baum et al., 2005] or stratocumulus clouds (ice) [Hale and Querry, 1973], as described in Meadows et al. [2018]. However, cloud optical depths as a function of pressure are derived from ROCKE-3D GCM output. For each pixel, there is a fixed solar zenith angle that is calculated from the star's sky position and that pixel's normal vector as seen in Figure 2. However, we run SMART over a hemispherical grid of observer zenith angles (defined by the radiative transfer streams) and observer azimuth angles, which are later interpolated to the observer's exact position (red vector in Figure 2). We use 16 streams (8 upwelling), which make up our 8 observer zenith angles, and we choose 7 observer azimuth angles uniformly spaced between 0-180 degrees. High-resolution UV through far IR spectra are simulated from 300 $cm^{-1}$ (~33 $\mu m$) to 100,000 $cm^{-1}$ (0.1 $\mu m$), with a wavenumber resolution of 1 $cm^{-1}$. Radiances are calculated for solar and thermal sources separately so that nightside pixels may contribute thermal emission to the disk-integrated spectrum while visible to the observer, but not illuminated by the star. The final results (stellar spectrum, wavelength grid, and radiance grids) for each pixel are then consolidated and compressed into a single HDF5 file for ease of access and use.

2.2.3. Production of Disk-Integrated Spectra

Given a sub-observer and sub-stellar points on the ROCKE-3D latitude-longitude grid, the SMART radiance database from the previous module is processed to produce a disk-integrated planet flux as it would be seen by a distant exoplanet observer with a direct imaging telescope. We set the sub-stellar latitude and longitude based upon the illumination pattern of the tidally locked planet that includes obliquity (365x) because the sub-solar point in this case does not move with respect to longitude. The sub-observer longitude and latitude are user-specified quantities that are selected depending on the specific experiment under consideration, and control desired planet phase angle. With all angles specified, the gridded radiance field for each pixel is linearly interpolated to the observer position on the sky plane. We note that at lower resolutions (NSIDE ≤ 4), HEALPix pixels are large enough that many of the pixels will be partially illuminated and/or visible to the observer. This can cause pixels near the planet's terminator/limb to have a solar/observer zenith angle greater than 90° while still being partially illuminated/visible, resulting in an underestimation of the number of pixels that contribute to the disk-integrated spectrum. . To address this issue , we employ a HEALPix "sub-grid" which splits each (macro)pixel into 1,024 sub-pixels. An example of sub-gridding is seen in Figure 2. We compute the same geometric values for each of these sub-pixels as we do for the larger HEALPix pixels, allowing us to determine what fraction of each pixel is visible and/or illuminated. Finally, the disk-integrated spectrum is produced by summing the HEALpix radiances weighted by the illumination and visibility in the following equation:

$$f_{disk\ int.}(\lambda) = \sum_{i=1}^{N_{HPIX}} A_i\ (\hat{p}\cdot\hat{a}_i)\ S_i(\lambda)\ + B_i\ (\hat{p}\cdot\hat{b}_i)\ T_i(\lambda)$$

Here, $A_i$ is the fraction of sub-pixels in pixel i that are both illuminated and visible to the observer (orange sub-pixels in Figure 2), and $B_i$ is the fraction of sub-pixels in pixel i that are visible to the observer (orange and red sub-pixels in Figure 2). $\hat{p}$ is the observer's position vector in relation to the HEALPix grid. $\hat{a}_i$ is the vector describing the illuminated and visible area of pixel i, and $\hat{b}_i$ is the vector describing the visible area of pixel i. $S_i(\lambda)$ and $T_i(\lambda)$ are the spectra produced by interpolating the reflectance and emission radiance fields, respectively, for pixel i to the observer's position. The result is that we weight both illumination and viewing geometries in the calculation of a pixel's contribution to the disk-integrated spectrum.

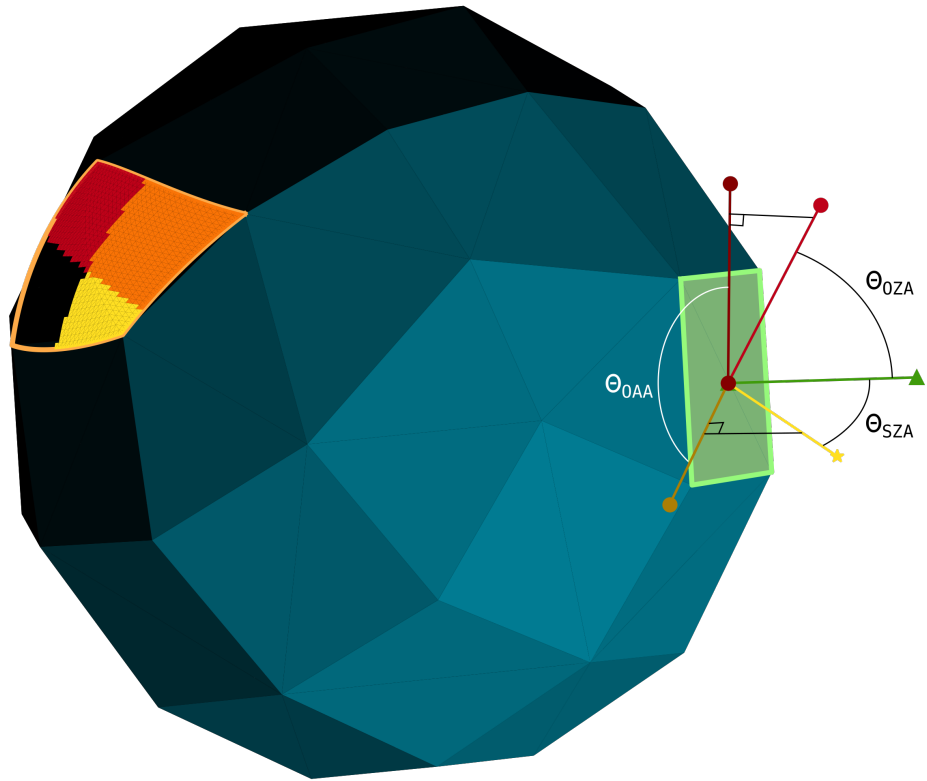

Figure 2. Illustrative example of the Hierarchical Equal Area isoLatitude Pixelization (HEALPix) grid used in this study and the geometric quantities involved. This HEALPix grid consists of 48 pixels, the same resolution our test cases were run with. Pixels that are more blue are closer to the subsolar point. The pixel highlighted in green shows the various angles and vectors used in translation from GCM outputs to SMART inputs. The green, red, and yellow vectors are the normal, observer position, and star position vectors, respectively, while the dark red and dark yellow vectors are the observer and star vectors, respectively, projected onto the pixel plane. These vectors are used to calculate the solar zenith angle (SZA), observer zenith angle (OZA), and observer azimuth angle (OAA) for each pixel. The orange highlighted pixel shows an example of the "sub-gridding" method employed here. The yellow sub-pixels are illuminated by the star, while the red sub-pixels are visible to the observer. Orange sub-pixels satisfy both these conditions, and contribute to the calculation of both visible, illuminated, and visible + illuminated fractions of a given pixel.

# 3. RESULTS

## 3.1. Simulated Climate and Atmospheric Dynamics

The influence of planetary rotation rate on climate has been well-known for decades [Williams and Holloway, 1982; Del Genio and Suozzo, 1987; Navarra and Boccaletti, 2002; Showman et al., 2015] as it naturally flows from the foundational equations of fluid motion on a rotating planet, termed the "primitive equations" by Andrews, Holton, and Leovy [1987], amongst others. That rotation rate, often represented as $\Omega$, is directly included in the Coriolis parameter and through the centrifugal forces imparted to the fluid by the rotating planet. In our simulations, we have slowed the rotation rate of the planet and hence reduced $\Omega$ by the integer factors shown in Table 1. This equivalently reduces the Coriolis parameter and centrifugal force for our simulated planets. The impact of this reduced Coriolis force is well-known as it broadens the overturning Hadley circulation (Figure 4 ), changes and eventually eliminates the characteristic westerly jet streams (Figure 6 ) of Earth (and Mars) and associated transient weather systems (baroclinic waves), and hence feeds into the pattern of clouds and precipitation around the planet (Figure 3 ). It is this later factor that is particularly relevant to the observation of exoplanets by large space telescopes such as Origins, LUVOIR, or HabEx and the habitability of terrestrial exoplanets (e.g., Yang et al., 2013; 2014b; Way et al., 2016) as it modifies the reflected and thermal light spectra of the planet (see Sections 3.2, 3.3, and 4).

Our focus in this section is to present the climate of our simulated planets to both provide a point of reference for our subsequent discussion on how the reflectance and thermal spectra are changed and to demonstrate that ROCKE-3D produces the expected changes to planetary climate due to a slowed rotation rate. Way et al. [2018] provides a more comprehensive description of the influence of rotation rate (and other factors) on the climate and hence habitability of terrestrial planets as simulated in ROCKE-3D.

Following Kopparapu et al. [2017], Figure 3 presents the surface air temperature, planetary albedo, column-integrated cloud condensed water, and total cloud cover for each simulation. Each plot in Figure 3 represents the average of the last 10 years of the model simulation, except for the 256x simulation, where the average of the last 50 years is shown. 243x is not shown in Figure 3 as it is substantially similar to 256x.

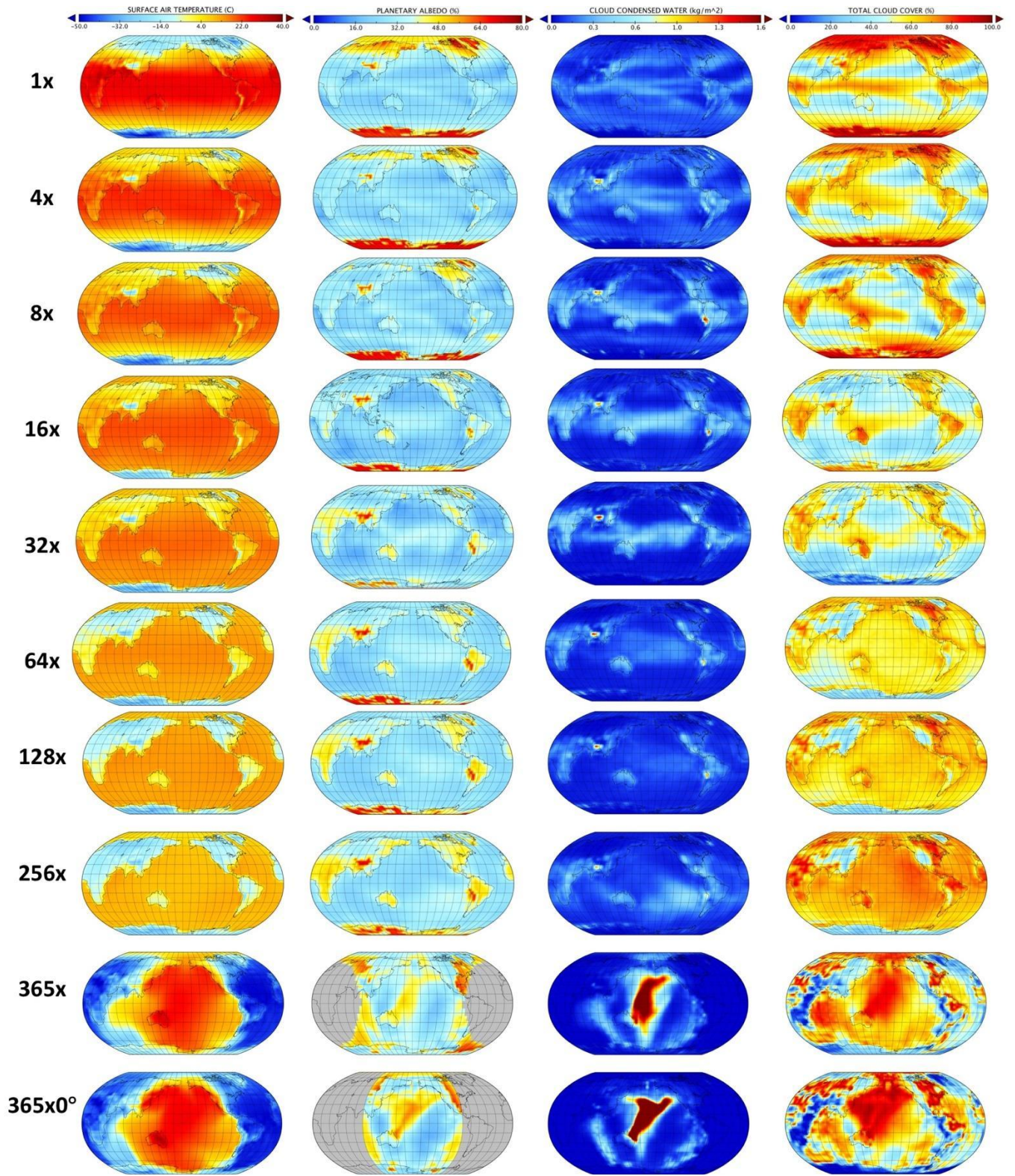

Figure 3. The surface air temperature (°C), planetary albedo (%), column-integrated cloud-condensed water (kg/m$^2$), and total cloud cover (%) from left-to-right for each specified ROCKE-3D GCM simulation. Each plot shown represents

an average of the last 10 years simulated, except for the 256x simulation which shows an average of the last 50 years.

As the rotation rate slows and the days lengthen, the planet initially cools (from a global mean surface temperature of 16.2°C for 1x to 11.6°C for 32x) while also becoming less cloudy. This is due to the broadening of the Hadley circulation (Figure 4) and a larger area of descending drier air in the mid-latitudes, resulting in reduced cloud cover and increased outgoing longwave radiation at middle and high latitudes. Indeed, between 8x and 16x day lengths, the Hadley cell becomes fully equator-to-pole (Figure 4) and the counter-rotating Ferrel cells are not present. Navarro and Boccaletti [2002] note that their simulations lost the presence of a Ferrel cell between a 72hr and 144hr day length. Kaspi and Showman [2015] lose indications of a Ferrel cell between their $\Omega_e/4$ and $\Omega_e/8$ simulations with otherwise Earth-like conditions (corresponding to our 4x and 8x simulations). Our ROCKE-3D simulations clearly show a Ferrel cell in the 4x (i.e., 96hr day length) simulation and even a hint of one in the 8x simulation before being entirely absent by 16x. This broad Hadley cell reduces the equator-to-pole temperature gradient, causing the tropics to cool while the poles warm. This is also reflected in the planetary albedo (2$^{nd}$ column from left in Figure 3  ) via darker (reduced albedo) in high latitudes due to diminished snow and ice cover (Figure 5).

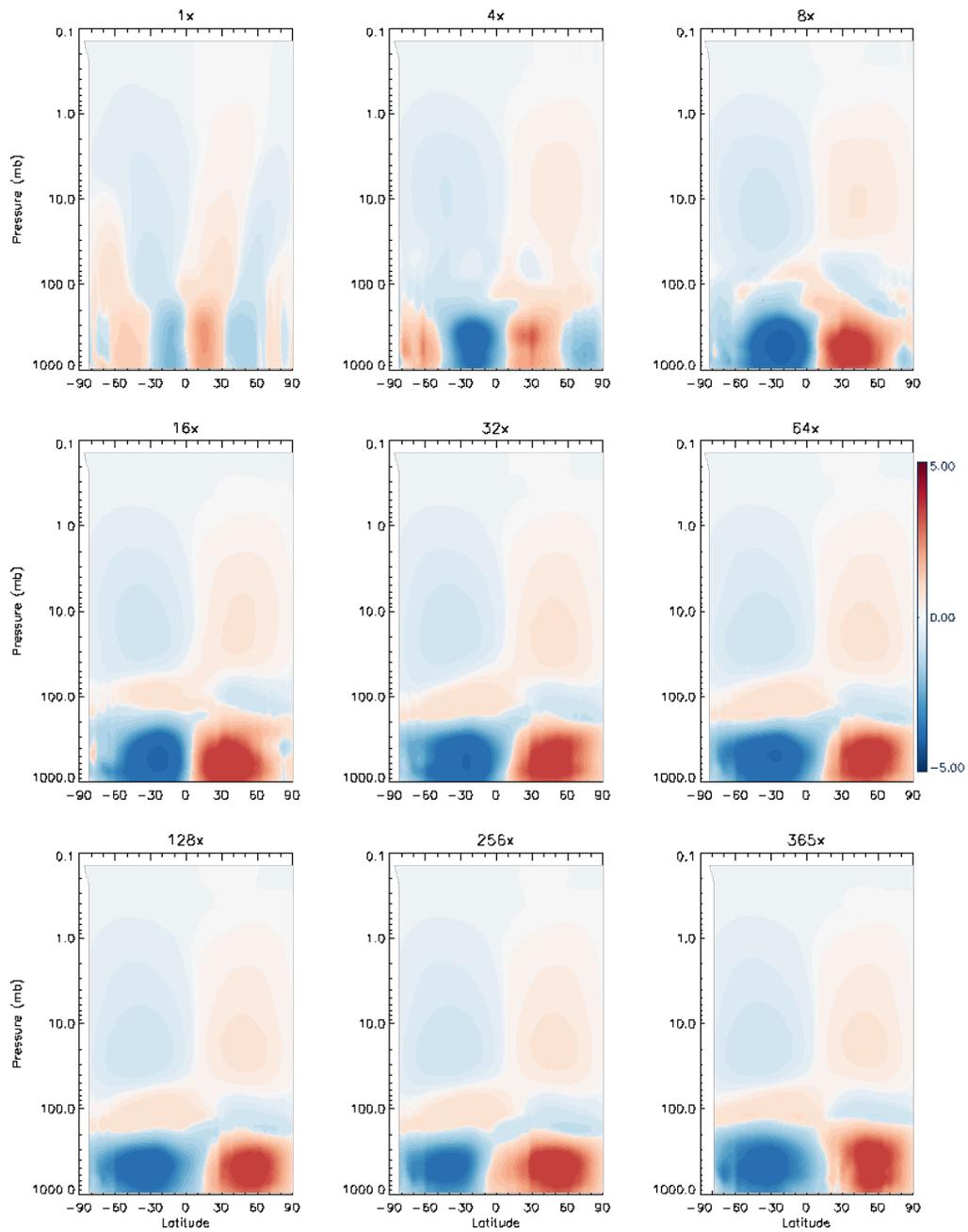

Figure 4. The Eulerian mean overturning circulation ($10^{10}$ kg/s) for 9 simulations from 1x (upper left) to 365x (lower right). Red (blue) colors indicate clockwise (counterclockwise) rotation.

Near a day length of 32x, cloud cover begins to expand around the planet and becomes increasingly predominant through the 256x simulation where most of the planet (save for specific isolated locations over continents) experiences cloud cover more than 70% of the time (Figure 3).  The 32x day length also marks a transition point between continental and ocean mean surface temperatures.  The increasingly weak near-surface wind field (not shown) with increased day length reduces the mixing of air between land surfaces and open ocean and the far larger heat capacity of the ocean keeps the ocean surface as much as 30°C warmer than the neighboring land surface, which cools much more than the ocean surface during the long night.  The expansion of continental snow cover and elimination of permanent sea ice is also of key importance in increasing the land-ocean temperature differences (Figure 5).  Even with a 4x day length, the Arctic ocean becomes permanently ice free (Figure 5) and most sea ice near Antarctica is eliminated as the mean surface temperature increases above freezing (Figure 3).  Land snow cover begins to noticeably expand again at a 32x day length, with the exception of Antarctica where snow cover declines at all increased rotation periods.  By 365x, as expected, the nightside of the planet (and part of the dayside near the terminator) is nearly entirely covered in ice and snow.  Note that we did not run our simulations to a point where the hydrological cycle reached equilibrium.  It is conceivable that over geological times, nearly all of the water on the dayside (i.e., in the Pacific Ocean and surrounding land surfaces) would be transported to the nightside as snow and ice and the dayside water reservoir would only be replenished by glacial flows returning to the dayside.  Such a scenario is discussed by Yang et al. [2014] (see also Menou, 2013; Leconte et al., 2013; Turbet et al., 2016), but ROCKE-3D both lacks a dynamic land ice model that accounts for the glacial flow that could possibly replenish the dayside with water (to some degree) and the duration of simulation time necessary to largely evacuate the Pacific Ocean is well beyond reasonable simulation run durations.

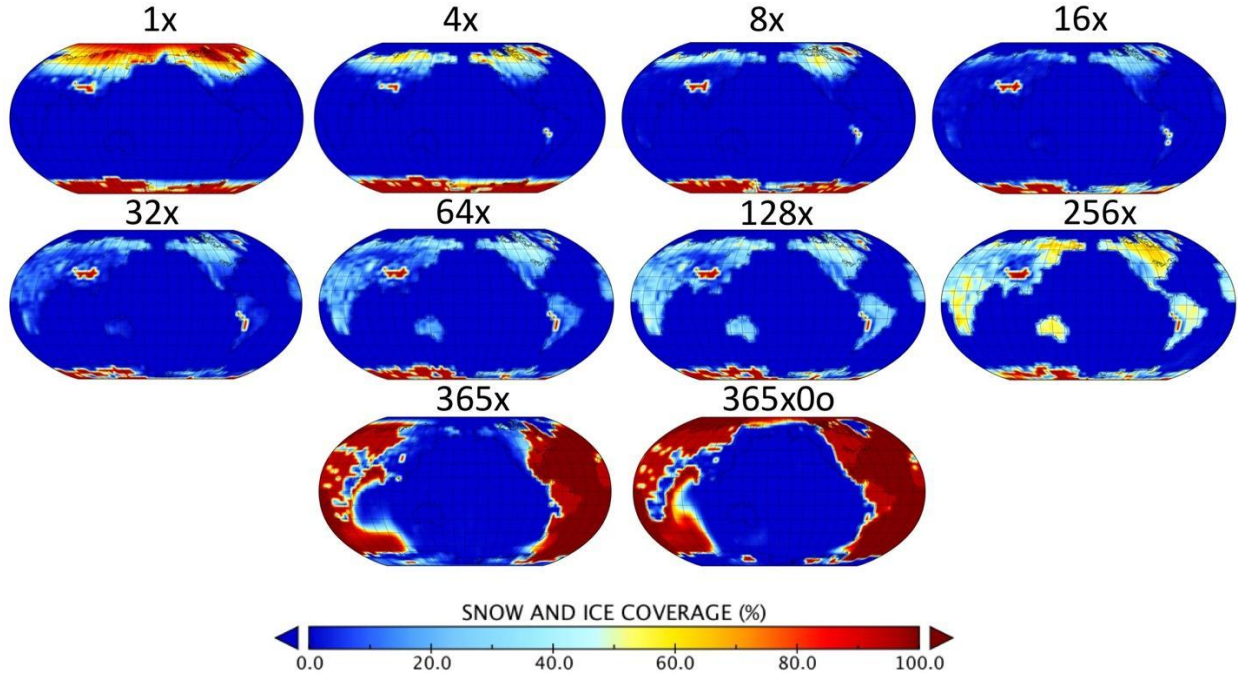

Figure 5. Snow and ice coverage (%) for 10 ROCKE-3D simulations.

The 243x and 256x simulations occur at or near (respectively) the 3:2 spin-orbit resonance for Earth. This results in unique climate impacts on yearly and decadal timescales where a small area of the planet (well less than 1 full hemisphere as in the 365x simulations) is in permanent night. As expected, snow and ice builds up near this naturally colder location, but this region of snow and ice cover moves eastward around the planet, particularly in the northern hemisphere, at slow timescales. This produces a regular "beat" to the climate (e.g., Figure 1) driven by the land and ocean distribution on the planet and the corresponding impact of relatively warmer or colder surface temperatures and snow and ice coverage as the small permanently shadowed portion of the planet moves across that land and ocean distribution.

The tidally-locked/sun-synchronous 365x simulations produce familiar patterns [Yang et al., 2013; 2014b; 2019a; 2019b; Kopparapu et al., 2017; Fauchez et al., 2019] of a very thick sub-stellar cloud, primarily composed of ice particles at high altitudes (Figure 3  ), that is not perfectly round but instead modulated by the presence of continents and a dynamical ocean transporting heat poleward.  There is an intense dayside-to-nightside temperature gradient (with a global mean surface temperature below freezing) that drives radial wind flow away from the subsolar point (not shown) and also transports moisture that falls in the form of snow (Figure 5  ).  Note that because of this, the nightside of the 365x simulations are far from cloud-free.

The atmospheric dynamics of our simulated planets manifest in many of the ways previously discussed in the literature. Namely, as the rotation rate slows and day length increases, the planet begins to exhibit an equatorial superrotating jet stream and the westerly jet streams in the mid-latitudes are diminished or absent entirely. Indeed, the flow in the stratosphere (particularly the middle and upper stratosphere) switches between easterly in the 1x simulation to westerly and superrotating even in the 4x simulation. Laraia and Schneider [2015] noted that equatorial upper tropospheric flow became westerly when Earth's rotation rate was slowed by a factor of 2 (i.e., a "2x" day length simulation). The vertical and latitudinal extent of the superrotating jet stream expands through the 32x simulation (Figure 6 ), the same point at which the tropospheric westerly jet streams at mid-latitudes are entirely absent. By the 64x simulation, the superrotating jet stream is absent, with only very weak westerly flow in the stratosphere and weak easterly flow in the troposphere (Figure 6). This pattern persists until the 365x day length where radial flow away from the subsolar point dominates the troposphere and the stratosphere is essentially stagnant. Interestingly, our 128x, 243x, and 256x show no superrotation while Del Genio and Zhou [1996] showed strong superrotation in these ranges and Venus exhibits very strong superrotation at a (retrograde) 243x day length [e.g., Read and Lebonnois, 2018]. It has been suggested that the high opacity of Venus' atmosphere at high altitudes may help drive the superrotation there due to enhanced wave and tide forcing [e.g., Lebonnois et al., 2010].

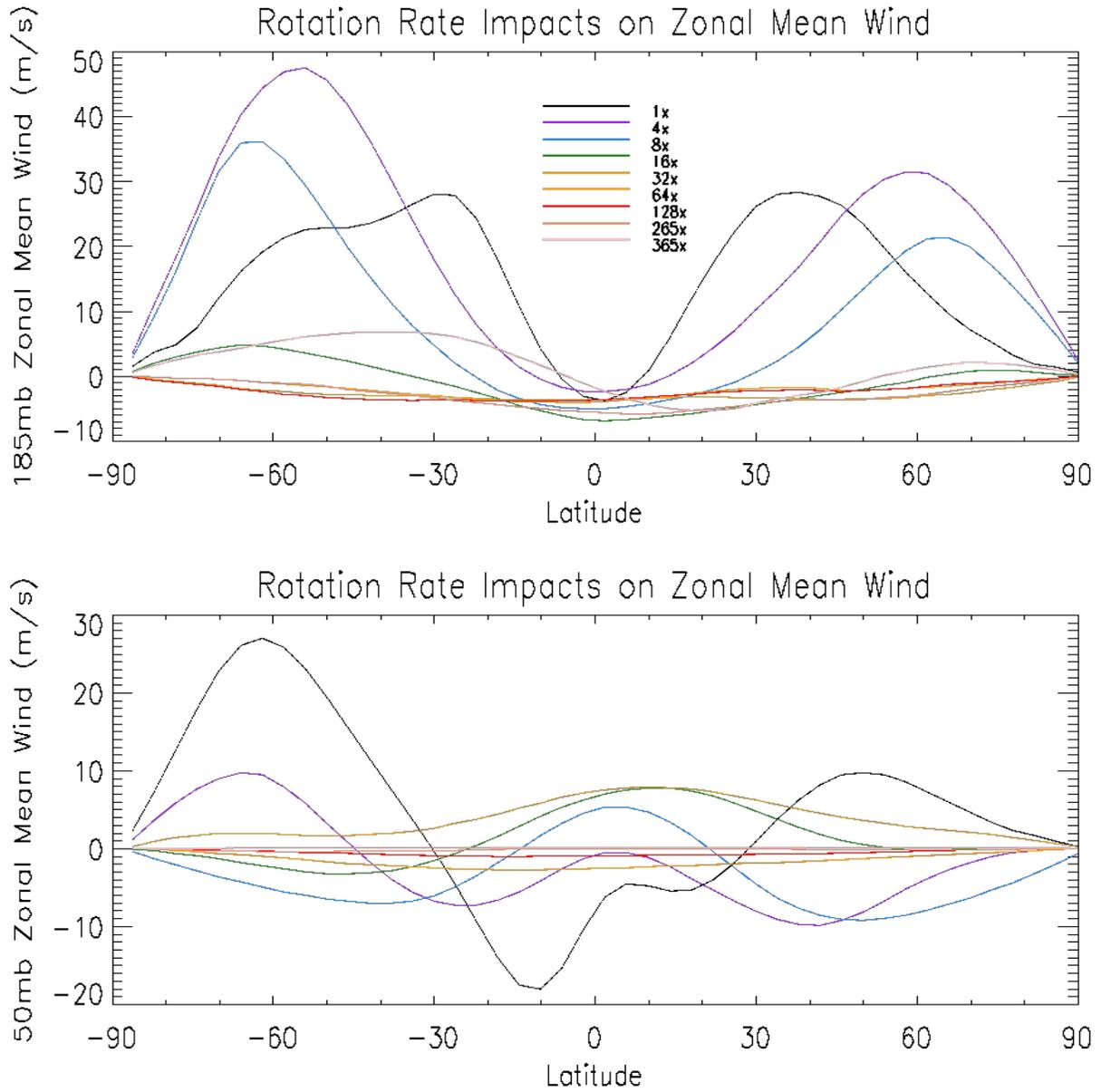

Figure 6. Zonal-mean zonal wind for the 185mb (top) and 50mb (bottom) pressure levels for most of the ROCKE-3D GCM simulations (see legend for corresponding line colors).

We touched on the changes to the tropospheric overturning circulation above, but we note that the stratospheric circulation also goes through a substantial change. Carone et al. [2018] describe an "Anti-Brewer-Dobson circulation" in the stratosphere of tidally-locked exoplanets and the same general pattern develops in our 8x and longer day length simulations. As seen in Figure 4 , 3 distinct overturning circulation cells form in the vertical column in each hemisphere, with generally an equator-to-pole Hadley circulation in the troposphere, a counter-rotating cell (i.e., pole-to-equator) near the tropopause, and then a second equator-to-pole cell in the stratosphere and mesosphere. As Carone et al. [2018] discussed, this likely has implications for photochemically-active species and their distributions in the atmospheres of slow rotating terrestrial planets, but it also may play a role in the escape of water from slow rotating planets. Specifically, the upper tropospheric/lower stratospheric tropical cold trap present on Earth is weakened for the slower rotating simulations, leading to a stratosphere that contains 30-70% more water vapor (e.g., ~0.003 g/kg at the 100 mb pressure level in the 32x simulation) relative to the Earth baseline (i.e., the 1x simulation with ~0.002 g/kg at the 100 mb pressure level) for the non-Sun-synchronous simulations and 100-200% more water vapor in the 365x simulations (not shown, ~0.0043 g/kg at the 100 mb pressure level).

## 3.2. Reflected Light Spectra

We computed the reflected light (visible and near-infrared) spectrum of the ROCKE-3D simulated planets using SMART to determine the observable changes driven by planetary rotation rate. Our discussion focuses on the 1x (Earth-like case), 64x, and 365x0° simulations as they are representative of fast rotators, slow rotators, and tidally-locked planets, respectively. Our analysis assumes a "face-on" orbital plane from the observer's point-of-view to simplify the dimensionality of our presented results. However, we note that variations in planet phase/inclination can also affect the observable reflectivity spectrum [e.g., Cahoy et al., 2010; Nayak et al., 2017], which may be degenerate with our spectral results.

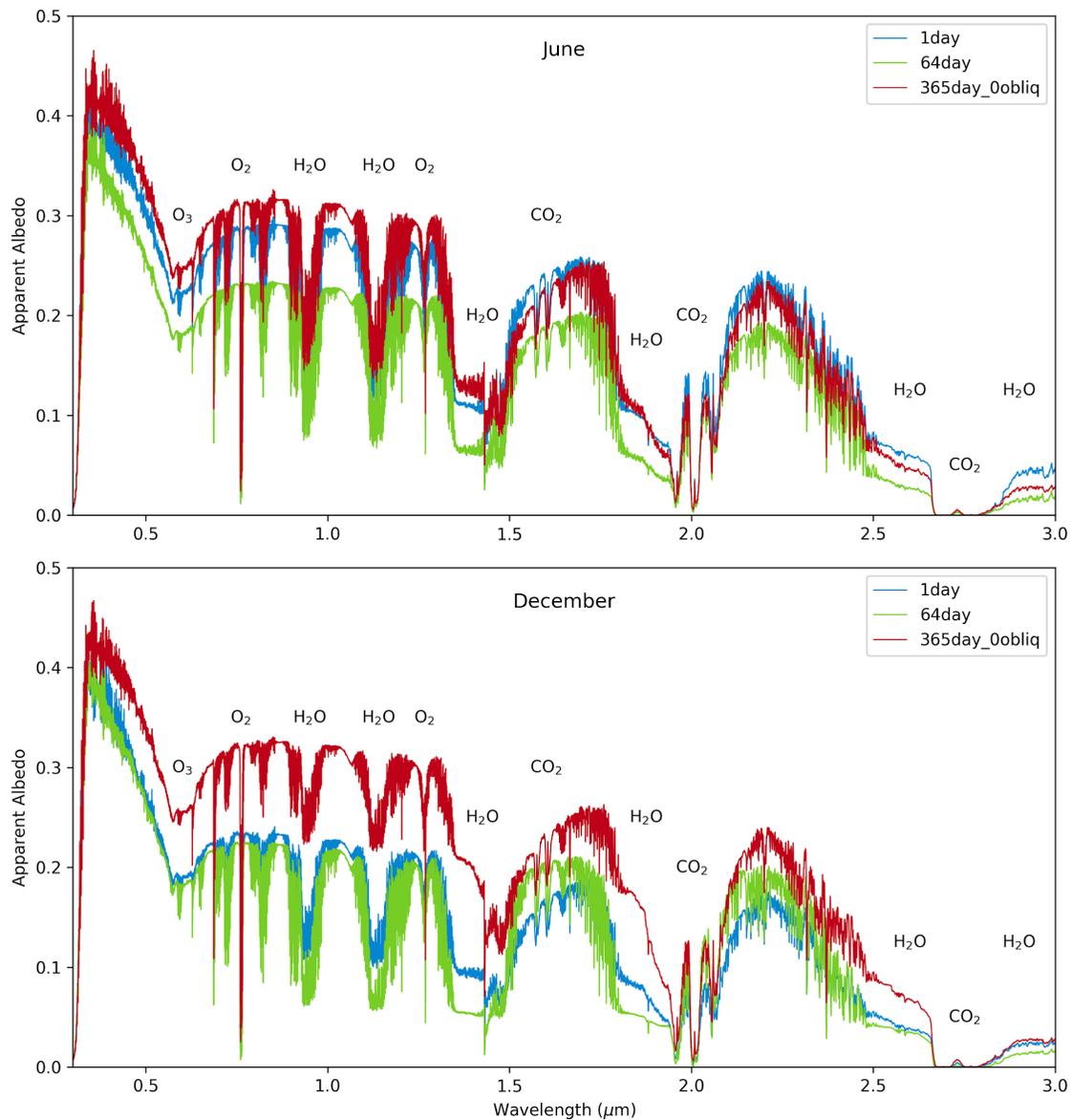

Figure 7. Comparison of albedo spectra for the 1x (blue), 64x (green), and 365x0°  (red) at the solstice months, June (northern summer; upper panel) and December (northern winter; lower panel). The observer is looking face-on to the orbital plane, towards the northern hemisphere. **Slower rotation rates correlate to less seasonal variation in these cases, and the synchronously rotating planet has a higher continuum albedo.**

Figure 7 shows reflected-light spectra for 1-day, 64-day, and 365-day rotation cases in the months of June and December. The reflection spectra are shown as the "apparent albedo", which is the geometric albedo the planet would have if it were a Lambertian sphere and is calculated by dividing the phase-dependent reflectance spectrum of the planet by the albedo of a Lambertian-scattering sphere at the same phase as the planet. This attempts to remove the effect of phase by scaling the spectrum so it approximates the planet at full phase and therefore approximates the geometric albedo. Note that the phase-effects of cloud scattering still remains in the apparent albedo spectra, which is slightly elevated at quadrature compared to the geometric albedo due to cloud forward scattering, which increases the reflectivity over that of a diffusely scattering Lambertian sphere [Robinson et al., 2010].

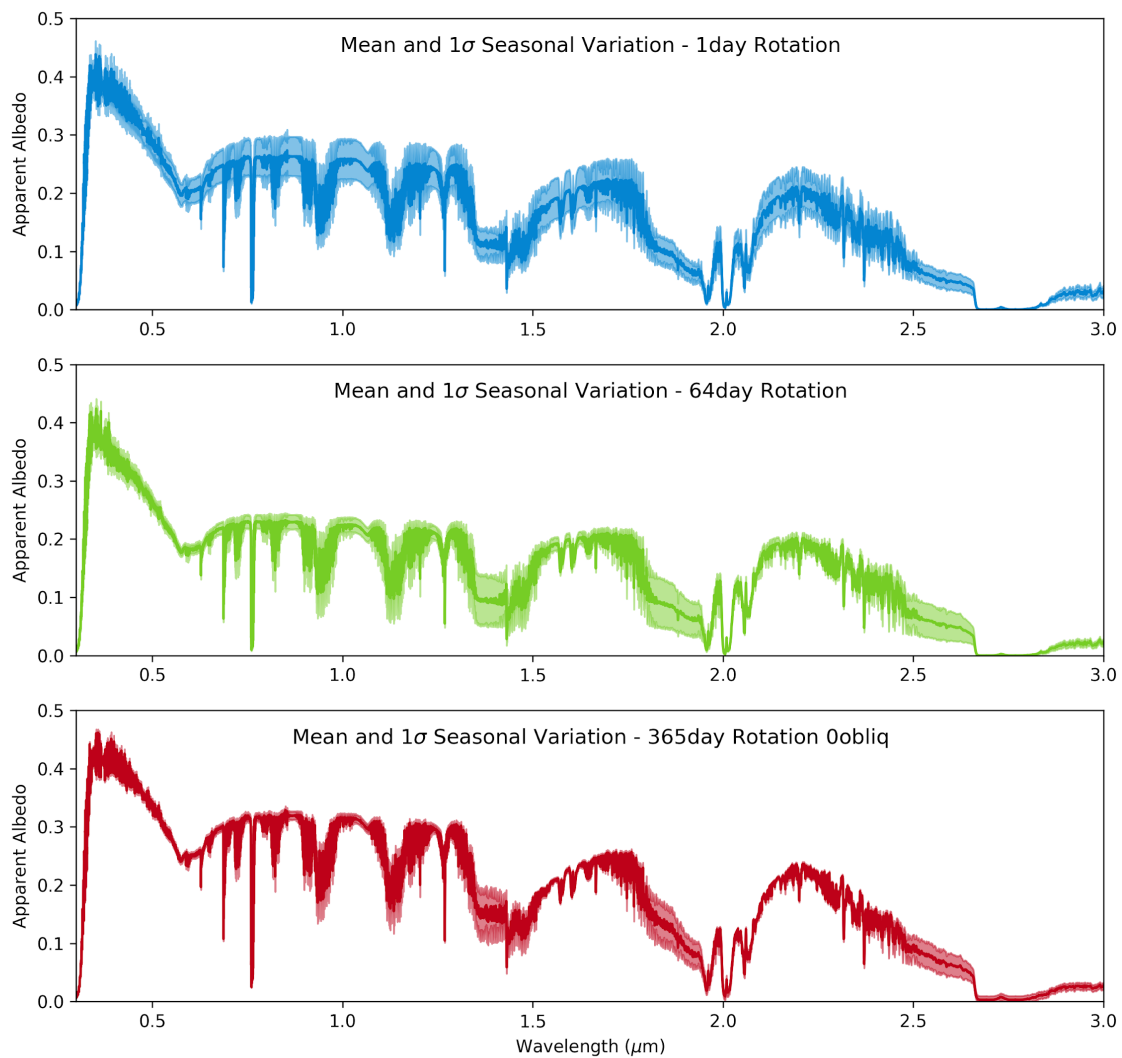

Figure 8. Comparison of yearly mean albedo spectra (solid lines) and 1σ monthly variations (semi-transparent envelopes) for the 1x (blue), 64x (green), and 365x0° (red) cases. The observer is looking face-on to the orbital plane, towards the northern hemisphere. Slower rotation rates correlate to less continuum variation throughout the year for these cases, and our synchronously

rotating planet exhibits little variation aside from the water features at 0.82μm, 0.94μm, 1.12μm, 1.4μm, 1.85μm, and 2.65μm.

In both Figure 7 and Figure 8, we see differences in the planet albedo spectra as a function of season for a given rotation rate, and as a function of rotation rate for a given season. In Figure 7, there is a relatively large difference (about 0.1) in the apparent albedo continuum levels of the 1x case between June and December compared to the 64x and 365x0° (both about 0.02 change in apparent albedo), indicating that the spectral variance is larger for faster rotating planets in our sampled cases. The continuum albedo of the 365x0° planet consistently exceeds the faster rotating cases, and the water absorption bands (e.g. 0.82, 0.94, 1.12, 1.4, 1.85, 2.65 μm) are not as deep as the other faster rotation states. Additionally, the intermediate 64x case has deeper water absorption features than the 1x and 365x0° in both months. These observable effects are nearly all a result of cloud coverage, cloud optical thickness, and cloud vertical extent, which we discuss in greater detail in Section 4.

Figure 8 further visualizes the seasonal spectral changes of the 1x, 64x and 365x0° cases by depicting the mean spectrum and standard deviation over all 12 months for each case. Note that at 1μm, the 1σ deviations of our cases are 0.035 (or 13.5% coefficient of variation, or CV), 0.013 (6.0% CV), and 0.007 (2.4% CV) for the 1x, 64x, and 365x0° cases respectively. The diminishing seasonal variation in the continuum with slowing rotation rate is due to differences in cloud properties and cover as well as surface albedo, which we will discuss in Section 4.

### 3.3 Thermal Emission Spectra

The thermal infrared emission spectra is modeled using SMART in an edge-on observer angle configuration that is amenable to analysis of primary and secondary transits by an exoplanet. We again focus on the 1x, 64x, and 365x0° simulations to provide representative cases for terrestrial planet rotation rates.

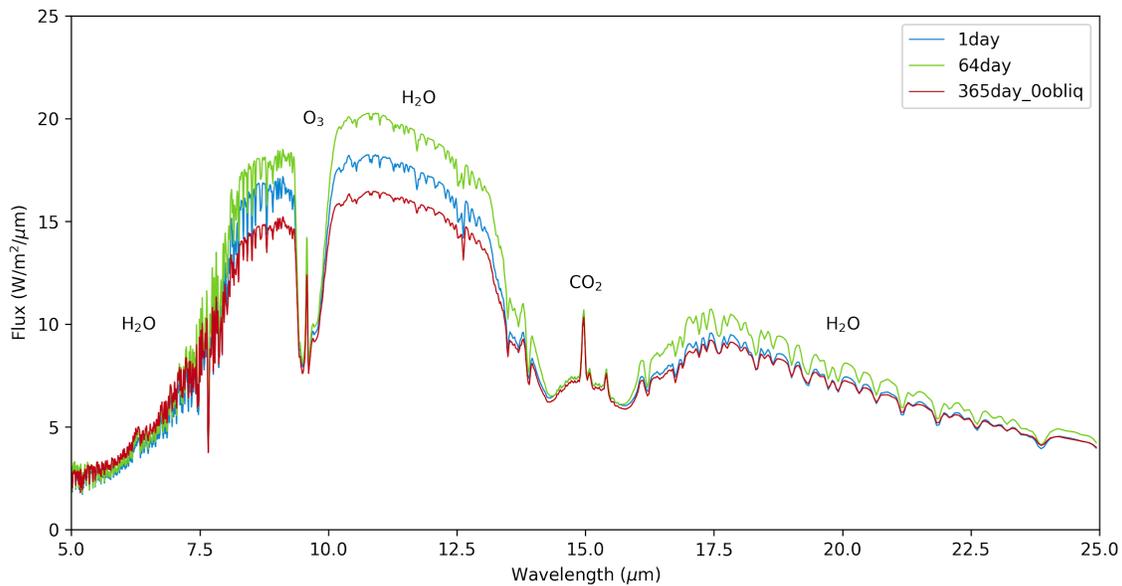

Figure 9. Thermal emission spectra of the 1x, 64x, and 365x0° cases at full phase in June (northern summer), viewed with the observer positioned on the ecliptic plane (edge-on inclination). These viewing angles are consistent with a secondary eclipse observation of a planet's dayside.

Figure 9 shows the thermal emission spectra of the 1x, 64x, and 365x0° rotation cases, viewed at full phase with an edge-on inclination. The most notable change is that the thermal continuum between 8µm and 13µm varies with rotation rate. The nominal Earth (1x) has a peak thermal flux of ~18 W/m$^2$/µm, the 64x peaks at ~20 W/m$^2$/µm, and the 365x0° peaks at ~16 W/m$^2$/µm, all at ~11 µm. The small absorption features present throughout this range are residual lines in the longwave tail of the 6.3 µm water band and the shortwave tail of the 50 µm water band. The prominent 15 µm CO2 band and the 9.6 µm O3 band both exhibit negligible change in the bottom of their bands with rotation rate for the cases examined. Like the Earth's true thermal emission spectrum, our models show an emission peak at 15 µm because temperature increases with altitude in the stratosphere, due to O$_3$ absorption. The central absorption peak of the 15 µm CO$_2$ band is so strongly absorbing that emission comes from near the top of the stratosphere, compared to the lower flux troughs surrounding the peak (at ~14 µm and 16 µm), which emit from near the top of the troposphere, which is colder than the top of the stratosphere. Removing O$_3$ or self-consistently modeling the photochemistry could change this behavior. However, the size of these absorption bands does vary between our rotation cases simply due to the aforementioned thermal continuum offsets.

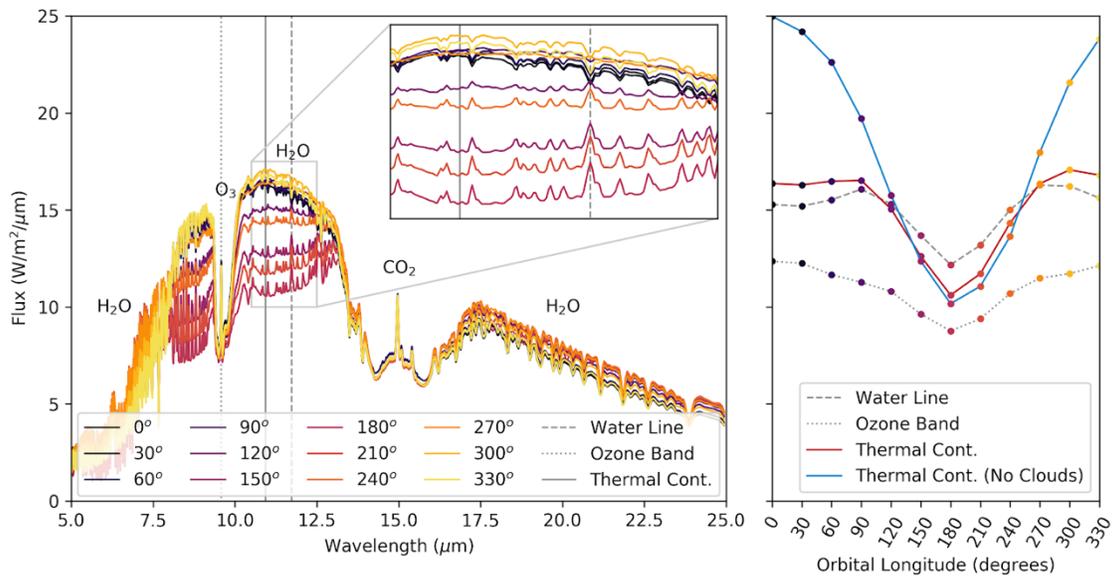

Figure 10. Thermal emission spectra (left) and lightcurves as a function of orbital phase sampled at various wavelengths (right) of the 365x0° case. Each spectrum is of the planet at various phases as an edge-on observer watches the planet throughout a full orbit, with 0° orbital longitude (dark purple) being full phase, 180° (magenta) being new phase, and 360° (yellow) being full phase again. The panel to the right shows the flux values as the planet moves through these phases at three different wavelengths; the dashed line samples a $H_2O$ line at 11.7 μm, the dotted line samples the $O_3$ band at 9.6 μm, and the solid lines sample the thermal continuum at 10.9 μm. The red solid line samples the thermal continuum with clouds included in the radiative transfer calculations, while the blue solid line samples the thermal continuum with clouds excluded from the radiative transfer calculations. Additionally, these wavelengths are plotted as vertical lines with their respective line styles in the left panel. Subtle changes with phase to the MIR thermal continuum and water absorption features may indicate large-scale dayside cloud structures that are consistent with GCM predictions for synchronously-rotating planets.

Figure 10 shows the thermal emission from the tidally locked and synchronously rotating planet as the planet orbits around the Sun as seen from an edge-on inclination observer. As the synchronously rotating Earth orbits the Sun, the edge-on observer sees different sub-observer longitudes at different planet phases, which may be sensitive to different depths into the atmosphere , due to the longitudinal distribution of clouds. The planet's thermal continuum peaks at full phase (light red or dark red) when the dayside of the planet is in view, and is at its minimum at new phase (bright red) when the nightside of the planet is in view. The maximum flux of the thermal continuum (solid red line in the right panel of Figure 10) is seen to be much lower than the maximum that occurs when we exclude clouds from the radiative transfer calculations (solid blue line in the right panel of Figure 10 ), but the minima of these two cases are much closer. This indicates that clouds are effective at muting thermal emission from the dayside, but they negligibly affect emission from the nightside for our synchronously rotating Earth case. Both the $O_3$ band at 9.6 µm and the $H_2O$ lines near 11.75 µm show similar sinusoidal trends with orbital longitude. However, the water lines transition from absorption to emission as the planet moves from full phase to new phase. This can be seen in the zoomed inset axis in the left panel of Figure 10. The $CO_2$ band at 15µm, which is not one of the sampled wavelengths, exhibits negligible change throughout phase. The phase-resolved thermal flux curves in Figure 10 are shown for individual wavelength resolution elements from our line-by-line radiative transfer calculations, which correspond to a spectral resolving power of R~900 near 11 µm. The water lines in this spectral region that exhibit transitions with phase from absorption to emission would require a spectrograph with at least R=200 at 11 µm to be individually resolved by a single spectral resolution element.

## 4. DISCUSSION

### 4.1. Climate

As discussed in the introduction, rotation rate can affect the habitability of a planet through changes in the global mean surface temperatures [Way et al. 2018] and carbonate-silicate weathering process [Jansen et al. 2019]. Our work also highlights this aspect, as shown in Figs. 3, 4 and 5. Our ROCKE-3D simulations recreate the expected climate behavior for slowing planetary rotation rate seen in previous theoretical and modeling work [Williams and Holloway, 1982; Del Genio and Suozzo, 1987; Navarra and Boccaletti, 2002; Showman et al., 2015; Way et al., 2018]. Specifically, the equator-to-pole temperature gradient weakens with increased day length/slower rotation rates due to an expanded Hadley circulation and the eventual elimination of the Ferrel cells between 8x and 16x day lengths. The climate cools between 1x and 32x day lengths and steadily becomes cloudier at longer day lengths/slower rotation speeds as the planet becomes an "all-tropics" planet from a dynamical regime. Sea ice is reduced from the Earth baseline in all cases until a tidally-locked/Sun-synchronous rotation rate is reached and a permanent nightside allows the ocean to freeze. Land snow and ice coverage decreases with increasing day length until temperatures become sufficiently cold (and nights sufficiently long) to allow snow to cover all continental areas seasonally/diurnally. These changes to cloud cover and snow/ice cover are of particular importance when evaluating how rotation rate drives changes to a planet's reflectance or thermal emission spectrum.

## 4.2. Reflected Light Spectra

Figure 7 illustrate how some of the climatic phenomena discussed above are revealed in the reflected light spectra of these planets. In particular, the slower rotating planets in our sample tend to exhibit less month-to-month variability in their reflected light spectra compared with our sample's faster rotators. The 1x case's continuum albedo spectrum is higher in June than in December, in part due to the visibility of the northern ice cap. In June, the ice cap is smallest but Earth's obliquity tilts the ice cap towards the sun where it reflects strongly in the visible. In December, even though the northern ice cap is at its largest it is tilted away from the sun, making most of the reflecting surface continent and ocean, both of which are much less reflective than ice in the visible. This trend is also seen for the 64x case, though the decrease is less pronounced. However, because the 365x0° lacks obliquity and is tidally locked, the surface of the planet and cloud coverage, and therefore its spectral features, are seen to experience negligible seasonal change.

The effect of these seasonal changes in surface albedo, or lack thereof, can also be seen in Figure 8. Sampling the continuum of the apparent albedo spectra at 1μm over all 12 months in each of our simulations, we observe that apparent albedo spectra values are 0.257±0.034 (13.4% coefficient of variation, or CV), 0.224±0.013 (5.6%CV), and 0.314±0.007 (2.3%CV) for the 1x, 64x, and 365x0° cases, respectively. This decrease in continuum variability with decreasing rotation rate is the result of seasonal changes in planetary surface features, like the northern ice cap, as discussed above.

There are also changes in the depth of absorption features as a function of both planetary rotation and season. As can be seen in Figure 7, the 1x Earth rotation case water bands in the NIR at 1.4 μm and 1.8 μm are deeper in June than in December, and the top panel of Figure 8 shows this feature to have an apparent albedo value of 0.111±0.025 (22.4%CV). This is due to increased cloud coverage during northern winter. Increased cloud coverage obscures the lower atmosphere via scattering optical depth, which competes with the water absorption optical depth, and decreases the size of the water bands. In contrast, the 64x has less cloud coverage and optically thinner clouds in December than in June, allowing reflected light to probe deeper into the atmosphere where there is more water, producing the deeper water features we observe in December. The large variance of the 64x case's 1.4 μm water feature in the middle panel of Figure 8, with values of 0.094±0.042(45.1%CV), is due to these changes in both cloud coverage and cloud optical depth. While the synchronously rotating case with zero obliquity (365x0°) shows little seasonal change in the continuum, the water bands seen in Figure 7 are deeper in June than in December. This change in water absorption can also be seen in the bottom panel of Figure 8, where the water feature near 1.4μm has an apparent albedo value of 0.148±0.038 (25.6%CV) around 1.4μm in Figure 8. Cloud coverage in this case varies very little (5%CV) over the course of a full year, and the variance in water absorption of the 365x0° is due to differences in water and ice cloud optical depths across the planet's surface.

The $O_3$ Chappuis band between 0.4 - 0.65 μm and $O_2$ A-band (0.76 μm) appear more prominent in the reflected light spectrum in the presence of significant cloud coverage, which we see consistently in our simulations on the day side of the 365x planet throughout the entire year. This occurs because water clouds increase the continuum reflectivity in the optical, which increases the equivalent width of gaseous absorption features that occur at or above the altitude of the cloud deck (e.g. Rugheimer et al. 2013).

### 4.3. Thermal Emission Spectra

The relationship between rotation rate and observed thermal flux in Figure 9 is the result of many factors including atmospheric temperature structure, molecular composition, and surface temperature. However, we find that clouds exert more influence on the emerging spectra than any of the phenomena listed above. Because clouds reflect strongly in the visible and absorb and re-emit heavily in the NIR/IR, many spectral features that would arise from phenomena taking place below the cloud deck are not seen by the observer. Comparing the 1x (Modern Earth) and 64x, although the 1 x has less uniform cloud coverage than the 64x case (17.6%CV vs 10.2%CV across all pixels), the 1x water clouds extend higher into the troposphere (reach optical depth of $10^{-3}$ at 0.2 bar vs. 0.4 bar) and, at 0.8 bar where the clouds reach their maximum optical depth, are twice as optically thick. The result of this is that the observer sees deeper into the atmosphere, where it is hotter, in the 64x than in the 1x, and the thermal continuum is correspondingly higher. The 365x0° (synchronously rotating Earth) has high-altitude (reach optical depth of $10^{-3}$ at 0.2bar), very optically thick (10 at 0.8 bar) clouds covering most of the planets dayside, which truncate the atmosphere at a higher, colder altitude in the troposphere, and so the observer sees a lower thermal continuum. Clouds extend higher in the 1x case relative to the slower rotation cases (until 365x) due to the deeper Hadley Circulation (Figure 3) and a generally more stable troposphere in the slow-rotating planets.

The interplay between cloud altitude and the atmospheric vertical thermal/compositional structure also affects molecular absorption features in the thermal emission spectrum. Figure 9 shows numerous water absorption features throughout the mid-IR for all the rotation cases, but they are slightly stronger for the 1x (Modern Earth) and 64x than for the 365x0° (tidally locked Earth). This is mainly due to cloud coverage, where the clouds in the 365x0° are much more optically thick and extend higher in the atmosphere than for the 1x and 64x. These thick clouds control the altitude from which the thermal continuum emerges, which is above the bulk of the water column since the water abundance decreases with altitude. Even though the 365x0° case has a more humid stratosphere than the other rotation cases, the radiative effect of clouds dominates and, as a result, the water features are slightly weaker in the synchronous rotation case.

In Figure 9 we also see the effect of clouds on the 9.6 μm $O_3$ band. The bottom of this band is mostly independent of rotation rate, without the strengthened secondary emission core in the 64x. This is mainly due to the fact that, as a function of rotation rate, $O_3$ abundance in the stratosphere changes very little (<0.3% CV across a year for any given HEALPix pixel) (note our simulations do not include active photochemistry that could alter the $O_3$ distribution based on planetary rotation rate and potentially lead to larger variations in stratospheric gas abundances between the day and night sides of the planet; see Chen et al. 2018, 2019), and for all rotation cases the $O_3$ abundance peaks at altitudes above the clouds. Combining this with the minor fluctuations in the temperature of the stratosphere where $O_3$ is absorbing most heavily, the result is that the bottom of the $O_3$ band remains constant throughout our rotation cases. However, the depth of this band is still dependent on rotation rate due to the effect of clouds on the continuum as discussed above.

The result is that planets with low, optically thin clouds like the 64x planet have a higher thermal continuum, and therefore deeper $O_3$ bands than planets like the 365x0°, which have high altitude, optically thick clouds with a lower thermal continuum.

Interestingly, the effect of clouds leads to opposing trends in the detectability of $O_3$ as a function of rotation rate for reflected light versus thermal emission. As previously stated, in reflected light, the large dayside cloud on the synchronously rotating Earth strongly reflects light and effectively backlights ozone's Chappuis band, making it appear more prominent than the fiducial Earth. However in thermal emission, the large dayside cloud decreases the temperature contrast between the thermal continuum and bottom of the 9.6 $\mu$m $O_3$ band, making it appear less prominent than the fiducial Earth. Cloud particle properties control the scattering and absorption of light by the clouds, and could modify these effects if properties are substantially different than what the model produces. These properties are (and will likely remain) unconstrained on terrestrial exoplanets.

The thermal flux emitted by the synchronously rotating Earth (365x0°) shown in Figure 10 exhibits interesting trends in flux vs. orbital longitude (phase). In Figure 3 we see that the 365x0° has a large, high-altitude dayside cloud. As we noted for Figure 9, these clouds block outbound thermal radiation from the atmosphere below them, and therefore much of the flux we see on the dayside of the planet is emitted from this cloud layer at higher, colder altitudes in the troposphere. This is seen near the peak of the emission spectra near 0/360° (dark purple and yellow spectra) orbital longitude where the water features are weak and the continuum is at its maximum. As a note, the blue line in the right plot in Figure 10 samples the same wavelength and orbital space as the solid line, but clouds were removed from the radiative transfer calculations to assess their spectral effects (clouds were not removed from the GCM simulations). We see that the continuum is much higher for the clear-sky than for the cloudy case on the day-side of the planet (0/360° orbital longitude), further supporting that these clouds block emission from the lower troposphere. As the planet orbits from full phase to new phase (0° → 180° orbital longitude, or dark purple → magenta), the dayside and its large, subsolar cloud rotate out of view and the night side comes into view for the edge-on observer. Recall from Figure 5 that the nightside is covered almost entirely in ice due to snowfall caused by winds that carry hot, moist air from the dayside to the cold nightside. This cold surface temperature, along with a colder atmosphere due to complete lack of stellar insolation, corresponds to a large dip in the thermal continuum. Additionally, the observer begins receiving flux from lower in the atmosphere from the night side of the planet, which is significantly less cloudy than the day side. The fact that emission from the cold nightside of the planet emanates from deeper, warmer layers in the atmosphere, and emission from the hot dayside emanates from higher, cooler layers acts to decrease the amplitude of the planet's thermal phase curve, which is in agreement with previous findings for synchronously rotating planets (e.g. Yang et al., 2013).

In addition to clouds for the synchronously rotating planet reducing the day/night temperature contrast (or *inverting* it due to cloud behavior in the case of Yang et al., 2013) that would be seen in a phase curve, we find an interesting analog of this effect to the inversion of weak water lines in the mid-IR, which switch from absorption on the dayside, to emission on the nightside (Figure 10  ). As explained above, the water features at full phase are particularly weak, but as the nightside becomes visible, we are able to see deeper into the atmosphere. The temperature structure of the nightside of the tidally-locked Earth is morphologically similar to that of the Modern Earth, except that there is an additional low altitude temperature inversion, where the surface is much colder (~40 K) than the atmosphere directly above it until about 700 mbar, at which point the atmosphere begins to cool again until the tropopause. This phenomenon is due to circulation and transport of warm air from the dayside to the nightside, meaning water molecules at these mid-tropospheric altitudes are at a higher temperature than the air and surface below them, and so water is seen in emission. A similar phenomenon for water has been observed for the day/night difference on the Earth ( Prakash et. al. 2019, their Figure 4  ) , where the surface is hotter than the near-surface atmosphere in the day and the opposite is true at night. This effect is only seen near the peak of the thermal emission where the atmospheric opacity is low and thus the thermal continuum comes from the cold inversion layer near the surface. Longward of ~17.5 $\mu$m, the water opacity is higher such that the continuum emission probes above the near-surface inversion, and as a result water is only seen in absorption. At these longer wavelengths, the thermal phase curve does invert slightly (stronger emission from the nightside due to day-night cloud behavior), as seen in Yang et al. (2013).

## 4.4. Observability Considerations

The reflected-light spectra of slow rotating Earths has important implications for the search of habitable and inhabited exoplanets with next-generation direct-imaging missions, such as LUVOIR and HabEx (and possibly ground-based observations from large telescopes like the Extremely Large Telescope (Turbet et al., 2016; Boutle et al., 2017) and Very Large Telescope (Lovis et al., 2017)). Our finding that the synchronously rotating Earth case has a consistently higher continuum albedo than both the true Earth and intermediate slow rotation cases may help to better detect such planets and characterize them because (1) higher albedo planets will yield higher S/N observations in a fixed exposure time than lower albedo planets, and (2) the higher continuum albedo increases the contrast on many key spectral features such as for the biosignature gases $O_2$ and $O_3$. This same cloud-induced effect decreased the contrast on $H_2O$ bands by raising the bottom of the band, which could be a telltale sign that a directly imaged planet has thick dayside clouds and a cold trap, perhaps as a result of synchronous rotation. However, a planet with a high albedo surface and low water abundance may be degenerate with this interpretation. Feng et al. (2018) reported a retrieval degeneracy between surface albedo and cloud optical thickness, which may limit the unambiguous interpretation of cloudy atmospheres in reflected light. However, the extent to which wavelength-dependent information, such as aerosol scattering, surface albedo, and gaseous absorption may further exacerbate or help to break these degeneracies remains an open question for future works.

Additionally, muted seasonal changes to the albedo spectrum continuum seen in observations at different orbital epochs could indicate a slowly or synchronously rotating planet. Although, like spectral absorption features, such an interpretation of the spectral continuum may be degenerate with other factors that could also reduce observable seasonal variations. These factors may include planets with zero obliquity that may not have strong seasonal variations [e.g. Olson et al., 2018], planets inclined such that they are viewed equator-on where north-south hemispheres are averaged out [e.g. Schwieterman et al., 2018], and atmospheres with global aerosol coverage that entirely obscures the lower atmosphere.

We identified numerous interesting thermal emission effects for slowly rotating planets, but these are likely infeasible to observe for true Earth-Sun analogs due to the poor signal contrast of eclipse spectroscopy in this planet-to-star radius ratio regime. For example, the inverting water features discussed in the last paragraph of Section 4.2 exhibit a contrast flux signal of $10^{-2}$ ppm and require a spectral resolving power of R>200 at 11μm. However, if future observations of M-dwarf Earths show similar thermal emission features in secondary eclipse or phase curve measurements, they could potentially be explained by the observables of slowly rotating planets presented here. While dayside clouds on synchronously rotating planets in many ways increase the prospect of atmospheric characterization for planets imaged in reflected light, they can be a detriment in the thermal emission spectra. This is because clouds (1) lower the dayside thermal continuum flux and decrease the thermal phase curve amplitude, both of which would require longer exposure times to account for, and (2) this has a secondary effect of decreasing the contrast on mid-IR absorption features, such as those from $O_3$. Intriguingly, our result that water lines near 11 $\mu$m switch from absorption on the dayside to emission on the nightside could in principle be used in the future to identify a synchronously rotating planet orbiting an M dwarf in medium to high spectral resolution data (R>200) using the Doppler cross-correlation technique (R>25,000) in the mid-IR (e.g. Snellen et al., 2013, 2017; Brogi et al., 2014), phase-resolved emission spectroscopy [e.g. Stevenson et al., 2014], or phase-resolved infrared interferometry [e.g. Bracewell, 1976; Angel et al., 1986; Cockell et al., 2009; Defrère et al., 2018].

## 5. CONCLUSIONS

Planned and potential ground-based and space-based telescopes of the next two decades could make spectral analysis of terrestrial exoplanets around nearby G-type stars a reality. Successfully interpreting those observations requires a firm foundation of modeling to understand how the litany of factors that influence planetary climate (e.g., atmospheric gas mixtures, planet size and orbit, planetary obliquity and rotation rate, etc.) will drive spectral behavior in the reflected light or thermal emission spectra of those worlds. Of key importance will be to break (or constrain) the many degeneracies between those factors in hopes of uniquely constraining the climate of a terrestrial exoplanet.

In this work, we focused our efforts on understanding how planetary rotation rate (i.e., the length of the sidereal day) drove changes in the reflected and thermal spectra of otherwise Earth-like worlds. We simulated planets with day lengths from 1x (an Earth day) to 365x (Sun-synchronous) times the length of Earth's day using the ROCKE-3D GCM and then calculated their reflected light and thermal emission spectra using the SMART radiative transfer model.

Climates at slower rotation rates behaved largely as expected based on previous work with expanded Hadley circulations in the troposphere, cooler global temperatures, a reduction in sea ice, and an increase in cloud cover at day lengths slower than 32x (Figure 3 ). Of particular interest is the development of 3 unique circulation regimes in the vertical column (from the surface to the stratopause) at day lengths of 8x to 16x and longer (Figure 4 ). This pattern may be important for understanding the distribution of photochemically active species on terrestrial planets and may help regulate atmospheric water loss due to the substantial moistening (30% or more relative to Earth) of the stratosphere at rotation rates slower (day lengths longer) than Earth's.

The observational changes to the reflected light and thermal emission spectra of these simulated worlds were determined using SMART. In this work we discuss the 1x, 64x, and 365x0° (365x day length with zero obliquity, indicating a fully tidally-locked planet) as representative members of fast rotation, slow rotation, and tidally-locked regimes. We find that the presence and abundance of clouds is of critical importance to both the reflected light and thermal emission spectra and drives changes to many of the diagnostic spectral features used to assess habitability (e.g., $H_2O$, $O_2$, and $O_3$). For example, in the reflected light spectrum, the tidally-locked 365x0° world sees little seasonal variation to the spectrum and a higher continuum driven by the bright subsolar cloud on the dayside. These high clouds serve to limit the depth of near-infrared (and thermal infrared) $H_2O$ absorption lines, which are thus deeper for the 1x and 64x worlds where the effective observational surface is at a lower altitude due to the comparatively thinner clouds. Of note is a potentially diagnostic feature of synchronously rotating planets: mid-infrared $H_2O$ absorption lines that switch from absorption on the dayside to emission on the nightside due to the substantial change in atmospheric temperature between the dayside and nightside and the transport of water vapor away from the subsolar point.

## Acknowledgments

S.G., R.K.K., and M.J.W. acknowledge support from the GSFC Sellers Exoplanet Environments Collaboration (SEEC), which is funded in part by the NASA Planetary Science Division's Internal Scientist Funding Model. This work was performed as part of NASA's Virtual Planetary Laboratory, supported by the National Aeronautics and Space Administration through the NASA Astrobiology Institute under solicitation NNH12ZDA002C and Cooperative Agreement Number NNA13AA93A, and by the NASA Astrobiology Program under grant 80NSSC18K0829 as part of the Nexus for Exoplanet System Science (NExSS) research coordination network. The results reported herein benefitted from collaborations and information exchange within NExSS. This work made use of the advanced computational, storage, and networking infrastructure provided by the Hyak supercomputer system at the University of Washington. ROCKE-3D simulations were conducted via the NASA High-End Computing (HEC) Program through the NASA Center for Climate Simulation (NCCS) at Goddard Space Flight Center (http://www.nccs.nasa.gov).